\documentclass[nofootinbib,twocolumn,showpacs,preprintnumbers,pre,aps,superscriptaddress]{revtex4-1}

\usepackage[dvipdfmx]{graphicx}
\usepackage{bm}
\usepackage{amsmath}
\usepackage{amssymb}
\usepackage{color}

\begin{document}

\title{Odd Elasticity of a Catalytic Micromachine}

\author{Akira Kobayashi}

\affiliation{
Department of Chemistry, Graduate School of Science,
Tokyo Metropolitan University, Tokyo 192-0397, Japan}

\author{Kento Yasuda}

\affiliation{
Research Institute for Mathematical Sciences, 
Kyoto University, Kyoto 606-8502, Japan}

\author{Kenta Ishimoto}

\affiliation{
Research Institute for Mathematical Sciences, 
Kyoto University, Kyoto 606-8502, Japan}

\author{Li-Shing Lin}

\affiliation{
Department of Chemistry, Graduate School of Science,
Tokyo Metropolitan University, Tokyo 192-0397, Japan}

\author{Isamu Sou}

\affiliation{
Department of Chemistry, Graduate School of Science,
Tokyo Metropolitan University, Tokyo 192-0397, Japan}

\author{Yuto Hosaka}

\affiliation{
Max Planck Institute for Dynamics and Self-Organization (MPI DS), 
Am Fassberg 17, 37077 G\"{o}ttingen, Germany}

\author{Shigeyuki Komura}\email{komura@wiucas.ac.cn}

\affiliation{
Wenzhou Institute, University of Chinese Academy of Sciences, 
Wenzhou, Zhejiang 325001, China} 

\affiliation{
Oujiang Laboratory, Wenzhou, Zhejiang 325000, China}

\affiliation{
Department of Chemistry, Graduate School of Science,
Tokyo Metropolitan University, Tokyo 192-0397, Japan}


\begin{abstract}
We perform numerical simulations of a model micromachine driven by catalytic chemical reactions.
Our model includes a mechano-chemical coupling between the structural variables and the nonequilibrium 
variable describing the catalytic reactions.
The time-correlation functions of the structural variables are calculated and further analyzed in terms
of odd Langevin dynamics.
We obtain the effective odd elastic constant that manifests the broken time-reversal symmetry of a catalytic 
micromachine.
Within the simulation, we separately estimate the quantity called nonreciprocality and show that its
behavior is similar to that of the odd elasticity.
Our approach suggests a new method to extract the nonequilibrium properties of a micromachine only by 
measuring its structural dynamics.
\end{abstract}

\maketitle

\section{Introduction}
\label{Sec:Int}

In recent years, the physics of micromachines such as bacteria, motor proteins, and artificial molecular 
machines has been intensively studied~\cite{Toyabe15,Bechinger16,Brown20}.
A micromachine can be defined as a small object that extracts energy from chemical substances 
in the system and further exhibits mechanical functions~\cite{Dey16,HK22}.
The interplay between the structural dynamics of such a small object and associated chemical reaction 
is crucial for biological functions of a micromachine~\cite{Togashi10,Mugnai20,Hosaka20}.
Owing to the developments in nonequilibrium statistical mechanics and experimental techniques,
various research has been conducted to reveal the energetics of a single 
micromachine~\cite{Harada05,Toyabe10,Hayashi15,Ariga18}.

In our previous work, we proposed a model that describes cyclic state transitions of a micromachine driven by 
catalytic chemical reactions~\cite{Yasuda21a}.
As we shall explain later, 
we considered a mechano-chemical coupling of the variables representing the chemical reaction and the 
internal structural state of a micromachine.
To estimate the functionality of a micromachine, we focused on the physical quantity called ``nonreciprocality" 
that represents the area enclosed by a trajectory in the conformational space.
Such a quantity directly determines the average velocity of a three-sphere 
microswimmer~\cite{Golestanian08,Sou19,Sou21} or the crawling speed of a cell on a 
substrate~\cite{Tarama18,Leoni17}.

Recently, Scheibner \textit{et al.} introduced the concept of odd elasticity that is useful to characterize 
nonequilibrium active systems~\cite{Scheibner20,Fruchart22}. 
Odd elasticity arises from antisymmetric (odd) components of the elastic modulus tensor
that violates the energy conservation law and thus can exist only in active materials~\cite{Zhou20,Braverman21}
or biological systems~\cite{Tan22}. 
Recently, some of the present authors proposed a thermally driven microswimmer with odd elasticity and  
demonstrated that it can exhibit a directional locomotion~\cite{YHSK21,KYLSHK23}.
The concept of odd elasticity was also applied to Purcell's three-link swimmer model~\cite{Ishimoto22} 
and to robotic systems~\cite{Brandenbourger22}.
We also showed that antisymmetric parts of the time-correlation functions in odd Langevin systems are 
proportional to the odd elasticity~\cite{YIKLSHK22,YKLHSK22}.
Whilst many theoretical models are proposed to explain microscopic physical origins of odd elasticity, 
quantitative measurements of the odd elastic modulus are still lacking in these stochastic and nonequilibrium 
systems.

In this paper, we perform extensive numerical simulations of a model micromachine driven by catalytic 
chemical reactions~\cite{Yasuda21a}.
We calculate the time-correlation functions of the structural variables and analyze them in terms
of effective even and odd elasticity of coupled Langevin equations~\cite{YIKLSHK22}.
Importantly, the presence of odd elasticity reflects the broken time-reversal symmetry of a catalytic 
micromachine.
We also show that the nonreciprocality is proportional to the odd elasticity in the current 
stochastic model. 
Our approach provides us with a new method to extract the nonequilibrium properties of a micromachine
only by measuring its structural dynamics without invoking the dynamics of the chemical reaction~\cite{LYIHK22}.

The current work is based on our previous two papers in Refs.~\cite{Yasuda21a} and \cite{YIKLSHK22}, 
and further completes our general view on stochastic micromachines. 
In these works, we explicitly discuss the effects of thermal fluctuations which provide driving forces 
for micromachines.
One of the results specific to our stochastic system is that the average work is proportional to the 
square of the odd elastic constant (see Sec.~\ref{sec:Nonreciprocality})~\cite{YHSK21}.
It should be also emphasized that the concept of odd elasticity is not limited to elastic materials. 
Although we focus on catalytic micromachines, most of the results also apply to general odd stochastic 
systems such as the odd microswimmer~\cite{YIKLSHK22,YKLHSK22}.
In our previous paper in Ref.~\cite{Yasuda21a}, we solved our model almost analytically by using several 
approximations. 
In this work, on the other hand, we numerically solve the same equations in the presence of noise without
using any crude approximations.
Such a study is necessary for the complete understanding of the proposed model.

In  Sec.~\ref{Sec:Model}, we briefly review our model of a micromachine driven by a catalytic chemical 
reaction and explain the numerical method~\cite{Yasuda21a}. 
In Sec.~\ref{Sec:enzyme}, we show the results of the numerical simulations and discuss the 
properties of the time-correlation functions~\cite{YIKLSHK22}. 
Then we estimate effective even and odd elastic constants by using the result of the overdamped 
odd Langevin equations.
A summary of our work and some discussion are given in Sec.~\ref{Sec:Dis}.

\section{Catalytic micromachine}
\label{Sec:Model}

\subsection{Model}

\begin{figure}[tb]
\centering
\includegraphics[scale=0.5]{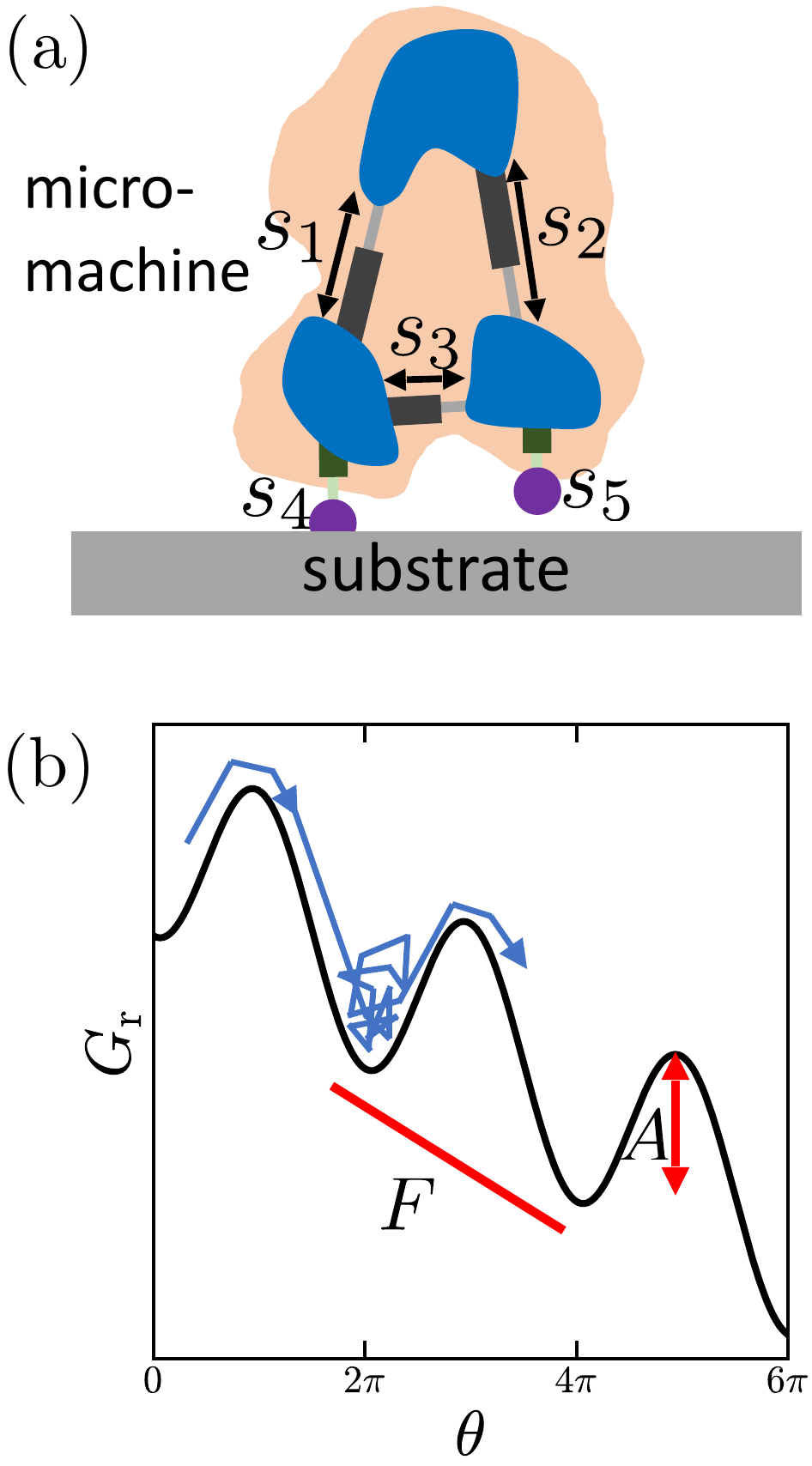}
\caption{(Color online)
(a) Schematic picture of a micromachine characterized by the conformational state variables $s_1$, $s_2$, and $s_3$.
Moreover, the adhesion between the domains and the substrate is described by the variables 
$s_4$ and $s_5$.
(b) The tilted periodic potential $G_\mathrm{r}(\theta)$, where $\theta$ is the catalytic reaction variable. 
As expressed in Eq.~(\ref{Gt}), $A$ is the energy barrier and $F$ corresponds to the nonequilibrium force.
A possible trajectory of $\theta$ is schematically shown by the blue arrow.
The value of $\theta$ fluctuates around the minimum of the potential and the transition to the next 
minimum takes place occasionally.
Reprinted figure with permission from Ref.~\cite{Yasuda21a}. 
Copyright (2021) by the American Physical Society.
}
\label{Fig:mod}
\end{figure}

We first explain our model of a catalytic micromachine which was introduced in 
Ref.~\cite{Yasuda21a} and is schematically shown in Fig.~\ref{Fig:mod}(a).
Consider a system which contains one enzyme molecule (E) that acts as a micromachine, 
substrate molecules (S), and product molecules (P).
The enzyme molecule plays the role of a catalyst and the corresponding chemical reaction is written 
as~\cite{Dillbook} 
\begin{align}
\mathrm{S}+\mathrm{E}\rightleftarrows \mathrm{ES}\to \mathrm{P}+\mathrm{E}
\end{align}
where ES indicates a complex molecule.
In our model, we introduce a dimensionless reaction variable $\theta(t)$ to quantify the extent of 
the catalytic reaction.
The reaction variable $\theta$ is a continuous number and increases $2\pi$ for each reaction.

According to the Kramers theory, the free energy $G_{\mathrm{r}}$ describing a chemical reaction is given 
by a tilted periodic potential~\cite{Hanggi90} 
\begin{align}
G_{\mathrm{r}}(\theta)=-A\cos\theta-F\theta,
\label{Gt}
\end{align}
as schematically represented in Fig.~\ref{Fig:mod}(b).
The first term is a periodic potential with a period of $2\pi$ and $A$ is the energy barrier.
This is because $\theta$ increases by $2\pi$ for one cycle of chemical reaction and should 
experience the same potential.
On the other hand, $F$ in Eq.~(\ref{Gt}) represents the chemical potential difference that drives 
catalytic reaction.
For example, $F$ represents the chemical potential change before and after ATP hydrolysis.

Next, we introduce the dimensionless state variables $s_i(t)$ ($i=1,2,3,\cdots$) characterizing the conformation 
of a micromachine.
As shown in Fig.~\ref{Fig:mod}(a), examples of the state variables are distances between the domains
in a micromachine.
To introduce the mechano-chemical coupling mechanism, we assume that each state variable $s_i$ 
experiences a harmonic potential, $(C_i/2)[s_i-\ell_i(\theta)]^2$, where $C_i$ is the coupling parameter 
and $\ell_i(\theta)$ is the dimensionless natural state that depends on the reaction variable $\theta$.
We consider that the natural state $\ell_i(\theta)$ changes periodically and assume 
the simplest periodic form $\ell_i (\theta)= \sin(\theta+\phi_i)$,
where $\phi_i$ is the constant phase difference relative to the reaction phase 
$\theta$~\cite{Mikhailov15,Canalejo21}.
Under these assumptions, we consider the following mechano-chemical coupling energy $G_\mathrm{c}$
between 
$\theta$ and $s_i$:
\begin{align}
G_\mathrm{c}(\theta,\{s_i\})=\sum_i\frac{C_i}{2}\left[ s_i- \sin(\theta+\phi_i)\right]^2.
\label{Gs}
\end{align}
Then the total free energy $G_{\mathrm{t}}$ in our model is simply given by 
\begin{align}
G_{\mathrm{t}}(\theta,\{s_i\})=G_\mathrm{r}(\theta)+G_\mathrm{c}(\theta,\{s_i\}).
\end{align}

For the time evolution of $\theta$ and $s_i$, we employ the Onsager's phenomenological 
equations in the presence of noises~\cite{KuboBook,DoiBook}
\begin{align}
\dot{\theta}&=-\mu_\theta\frac{\partial G_{\mathrm{t}}}{\partial\theta}+\xi_\theta(t),
\label{Dt1}\\
\dot{s_i}&=-\sum_{j}\mu_{ij}\frac{\partial G_{\mathrm{t}}}{\partial s_j}+ \xi_i(t),
\label{Ds1}
\end{align}
where dot indicates the time derivative.
In the above equations, the time-evolutions of the variables are proportional 
to the respective thermodynamic forces, and they relax to the thermal equilibrium state for which 
the total free energy is minimized.
Here, $\mu_\theta$ and $\mu_{ij}$ are the mobility coefficients, and $\xi_\theta$ and $\xi_i$ 
represent thermal fluctuations which satisfy the following 
fluctuation-dissipation theorem~\cite{KuboBook,DoiBook}
\begin{align}
&\langle\xi_\theta(t)\rangle=0, \quad \langle\xi_\theta(t)\xi_\theta(t')\rangle =2\mu_\theta k_\mathrm{B}T\delta(t-t'),\\
&\langle\xi_i(t)\rangle=0, \quad \langle \xi_i(t)\xi_j(t')\rangle =2 \mu_{ij}k_\mathrm{B}T\delta(t-t'), \\
&\langle \xi_\theta (t)\xi_i(t')\rangle = 0, 
\end{align}
where $k_{\rm B}$ is the Boltzmann constant and $T$ is the temperature.

Although our model is general and includes many degrees of freedom, we make several simplifications 
to capture the physical insight of a catalytic micromachine.
First, we only consider two degrees of freedom, i.e., $s_1$ and $s_2$.
Second, the mobility coefficients $\mu_{ij}$  ($i,j=1,2$) is assumed to have the form 
$\mu_{ij}=\mu_s \delta_{ij}$, where $\delta_{ij}$ is the Kronecker delta.
Third, the coupling free energy is symmetric between the two degrees of freedom, i.e., 
$C_1=C_2=C$. 
In this work, we investigate the case when $\phi_1=0$ because different choices
of $\phi_1$ give qualitatively the same $\phi_2$-dependence.
Then Eqs.~(\ref{Dt1}) and (\ref{Ds1}) reduce to the following simplified set of equations~\cite{Yasuda21a}
\begin{align}
\dot{ \theta}  = & -\mu_\theta \big[ A \sin \theta -F  
-C \cos \theta [ s_1- \sin \theta ] \nonumber\\
& -C \cos \left( \theta+ \phi_2 \right) [ s_2- \sin \left ( \theta + \phi_2 \right ) ] \big] +\xi_\theta,
\label{Dt2}\\
\dot{s_1}  = & -\mu_s C [ s_1- \sin \theta ] +\xi_1,
\label{Ds2a}\\
\dot{s_2}  = & -\mu_s C [ s_2- \sin \left ( \theta + \phi_2 \right ) ] +\xi_2.
\label{Ds2b}
\end{align}
Notice that the above equations are difficult to solve analytically because they 
are highly nonlinear although some attempts were made in Ref.~\cite{Yasuda21a}.

\subsection{Numerical method}

In the above model, the variables $\theta$ and $s_i$ are dimensionless, whereas $A$, $F$, and 
$C$ have the dimension of energy. 
The latter quantities are all scaled by the thermal energy $k_{\rm B}T$ and the corresponding dimensionless 
quantities are defined by $\hat{A}=A/(k_{\rm B}T)$, $\hat{F}=F/(k_{\rm B}T)$, and $\hat{C}=C/(k_{\rm B}T)$, 
respectively.
The dimensionless time is defined by $\hat{t}=2 \mu_\theta k_{\rm B}T t$ and the ratio between the two 
mobilities is fixed to $\mu_s/\mu_\theta=1$ in our simulation.

We numerically solve the coupled stochastic differential equations in Eqs.~(\ref{Dt2})--(\ref{Ds2b}) by 
integrating the following dimensionless quantities
\begin{align}
d \theta =  & (\hat{F} - \hat{A} \sin \theta ) d\hat{t}  
+\hat{C} \cos \theta [ s_1- \sin \theta ] d\hat{t} \nonumber\\
& + \hat{C} \cos \left( \theta+ \phi_2 \right) [ s_2- \sin \left ( \theta + \phi_2 \right ) ] d\hat{t}
+ dW_\theta,
\label{numDt2} \\
d s_1  = & - \hat{C} [ s_1- \sin \theta ] d\hat{t} + dW_1,
\label{numDs2a} \\
d s_2  = & - \hat{C} [ s_2- \sin \left ( \theta + \phi_2 \right ) ] d\hat{t} + dW_2,
\label{numDs2b}
\end{align}
where $dW_\theta, dW_1, dW_2 \sim \sqrt{dt}$ are the increment of Wiener processes.

\begin{figure}[tb]
\centering
\includegraphics[scale=0.45]{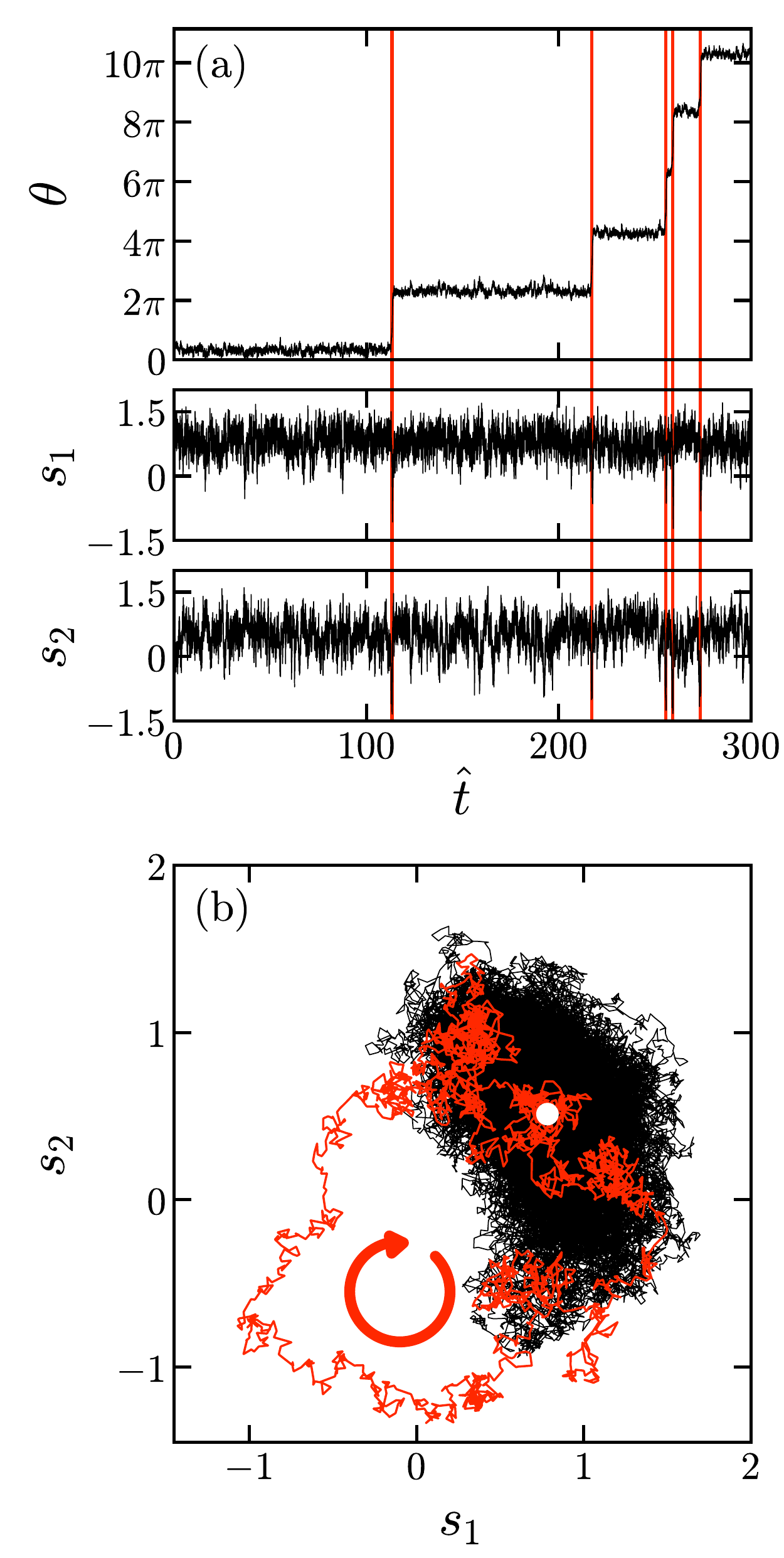}
\caption{(Color online)
(a) Time evolution of $\theta$, $s_1$, and $s_2$ as a function of dimensionless time 
$\hat{t}=2 \mu_\theta k_{\rm B}T t$ when 
$\hat{A}=20$, $\hat{F}=16$, $\hat{C}=20$, $\phi_2=\pi/2$, and $\mu_s/\mu_\theta=1$.
The red vertical lines indicate the moments when the catalytic chemical reactions take place.
(b) The trajectory of $s_1$ and $s_2$ corresponding to one cycle of chemical reaction that 
occurs in (a).
When $\theta$ fluctuates around the local minimum of the potential energy $G_{\rm r}(\theta)$, 
$s_1$ and $s_2$ also fluctuate (black trajectory for $115 \le \hat{t} < 216.5$) around the average 
position (white circle). 
When the chemical reaction takes place and $\theta$ changes by $2\pi$, the trajectory forms 
a closed loop (red trajectory for $112.5 \le  \hat{t} < 115$) in a clockwise manner as shown by the red 
circular arrow. 
}
\label{fig:rawdata}
\end{figure}

When we numerically calculate the statistical average $\langle X \rangle$ of a quantity $X$, we use 
the long-time average. 
This is justified because we have checked that the ensemble average and the time average give the 
same result in our simulation after the system reaches the steady state.

\section{Simulation results}
\label{Sec:enzyme}

\subsection{Time evolution}

As an example of the simulation result, we show in Fig.~\ref{fig:rawdata}(a) the time evolution of 
$\theta$ (top), $s_1$ (middle), and $s_2$ (bottom) for certain set of parameters. 
When the catalytic reaction occurs, we see that the reaction variable $\theta$ increases in a stepwise 
manner by $2\pi$, whereas the state variables $s_1$ and $s_2$ undergo almost random fluctuations. 
Corresponding to the occasional increase of $\theta$, both $s_1$ and $s_2$ tend to show 
peaks as indicated by the vertical red lines. 
Although the result in Fig.~\ref{fig:rawdata}(a) demonstrates that the present model 
micromachine is indeed driven by thermal fluctuations, it is difficult to isolate 
the peaks of $s_{i}$ because they are almost comparable to the background fluctuations.
Moreover, the state transitions become very rare when $\hat{A}\gg 1$.

In Fig.~\ref{fig:rawdata}(b), we plot a typical trajectory of $s_1$ and $s_2$ for one typical cycle of chemical reaction. 
While $\theta$ fluctuates around the local minimum,  $s_1$ and $s_2$ also fluctuate (black trajectory) around 
the average structure specified by the white circle.
When $\theta$ increases by $2\pi$, the trajectory of $s_1$ and $s_2$ follows a directional closed loop (red trajectory). 
The direction is clockwise when $0 < \phi_2 < \pi$ (as shown by the red arrow) or counterclockwise 
when $\pi < \phi_2 < 2\pi$.
The area enclosed by the trajectory can be obtained by the following quantity called 
``nonreciprocality"~\cite{Yasuda21a} 
\begin{align}
R_{12}=\oint dt\, \dot{s}_1 s_2,
\label{nonreciprocality}
\end{align}
where the integral is taken over one cycle. 
The statistical property of $R_{12}$ in our stochastic simulation will be discussed later in Sec.~\ref{sec:Nonreciprocality}. 
(We avoid to use the terminology ``nonreciprocity" that has broader and general 
meanings~\cite{Caloz18,Nassar20,Zhou22}.)

\subsection{Time-correlation functions}

\begin{figure}[tb]
\centering
\includegraphics[scale=0.45]{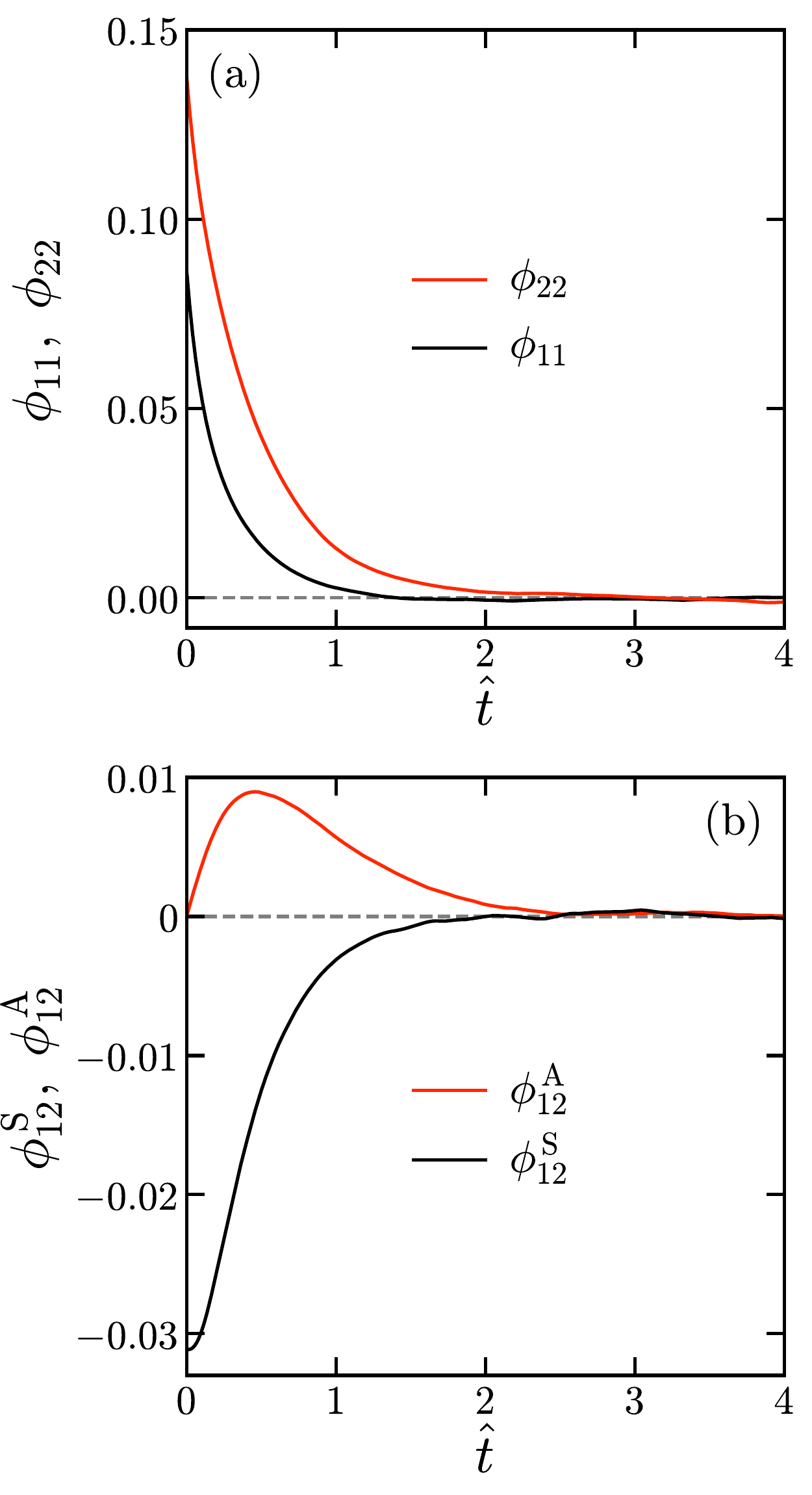}
\caption{(Color online)
(a) The self-correlation functions $\phi_{11}(t)$ (black) and $\phi_{22}(t)$ (red) as a function of 
the dimensionless time $\hat{t}$.
(b) The symmetric part of the cross correlation function $\phi_{12}^\mathrm{S}(t)=[\phi_{12}(t)+\phi_{21}(t)]/2$ 
(black) and the antisymmetric part of the cross correlation function 
$\phi_{12}^\mathrm{A}(t)=[\phi_{12}(t)-\phi_{21}(t)]/2$ (red) as a function of the dimensionless time $\hat{t}$.
For both (a) and (b), the parameters are the same as in Fig.~\ref{fig:rawdata}.
}
\label{fig:timecorrelation}
\end{figure}

From the time series of $s_1$ and $s_2$ in Fig.~\ref{fig:rawdata}(a), we calculate the structural 
time-correlation functions of a catalytic micromachine. 
We first define the time-dependent fluctuation by $x_i(t)=s_i (t)-\langle s_i \rangle$, 
where $\langle s_i \rangle$ is the average of $s_i$. 
Then the matrix of time-correlation functions $\phi_{ij}(t)$ is defined as    
\begin{align}
\phi_{ij}(t)=\langle x_i(t) x_j(0)\rangle=\phi_{ij}^\mathrm{S}(t)+\phi_{ij}^\mathrm{A}(t).
\end{align}
In the above, $\phi_{ij}(t)$ is decomposed into the symmetric and antisymmetric parts which 
satisfy $\phi_{ij}^\mathrm{S}(t)=\phi_{ji}^\mathrm{S}(t)$ and 
$\phi_{ij}^\mathrm{A}(t)=-\phi_{ji}^\mathrm{A}(t)$, respectively~\cite{YIKLSHK22}. 
For our later purpose, we also define the equal-time-correlation functions with a bar as 
$\bar{\phi}_{ij}=\phi_{ij}(0)$.

When the system is in equilibrium for which the time-reversal symmetry holds, the correlation 
functions must satisfy the reciprocal relation $\phi_{ij}^{\rm eq}(t)=\phi_{ji}^{\rm eq}(t)$ for 
$i \neq j$~\cite{KuboBook,DoiBook}.
This is equivalent to saying that the antisymmetric part of the correlation function 
$\phi_{ij}^\mathrm{A}(t)$ should vanish in equilibrium~\cite{YIKLSHK22}. 
In nonequilibrium situations, however, the antisymmetric parts can exist because the time-reversal 
symmetry can be generally violated~\cite{Epstein20,Hargus20,Han21}.

In Fig.~\ref{fig:timecorrelation}(a), we plot both the self-correlation functions $\phi_{11}(t)$ 
(black) and $\phi_{22}(t)$ (red) as a function of the dimensionless time $\hat{t}$. 
The overall behavior of the self-correlation functions is nearly described by an exponential function
with a slight modification.
Although the equal-time-correlation functions $\bar{\phi}_{11}$ and $\bar{\phi}_{22}$ take different values, 
the slopes of the initial decay are the same for these self-correlation functions.
Notice that the initial decay of the correlation function reflects the mobility of the system.

In Fig.~\ref{fig:timecorrelation}(b), we plot both the symmetric part of the cross correlation function 
$\phi_{12}^\mathrm{S}(t)=[\phi_{12}(t)+\phi_{21}(t)]/2$ (black) and the antisymmetric part of the 
cross correlation function $\phi_{12}^\mathrm{A}(t)=[\phi_{12}(t)-\phi_{21}(t)]/2$ (red) as a function 
of $\hat{t}$.
We clearly see that $\phi_{12}^\mathrm{A}(t)$ is nonzero, namely, $\phi_{12}(t) \neq \phi_{21}(t)$ for 
$0< \hat{t}<2$.
This means that the time-reversal symmetry is explicitly broken and hence the antisymmetric part 
of the cross-correlation function exists in the present catalytic system.
As we discuss below, the antisymmetric parts of the correlation functions can be quantitatively 
characterized by effective odd elasticity~\cite{YIKLSHK22}.

\begin{figure}[tb]
\centering
\includegraphics[scale=0.45]{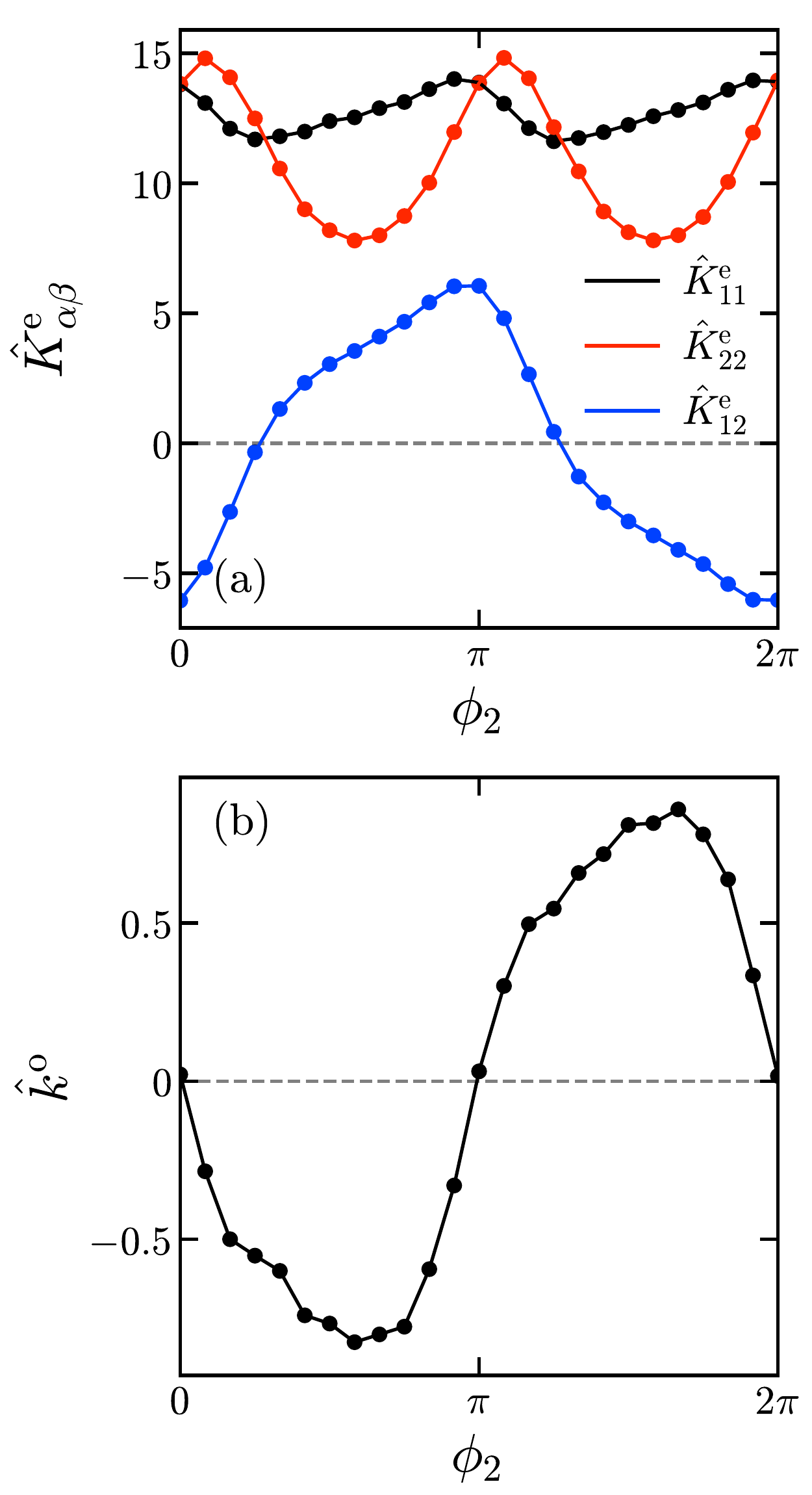}
\caption{(Color online)
(a) The dimensionless effective even elastic constants 
$\hat{K}_{11}^{\rm e}=K_{11}^{\rm e}/(k_{\rm B}T)$ (black), 
$\hat{K}_{22}^{\rm e}=K_{22}^{\rm e}/(k_{\rm B}T)$ (red), and 
$\hat{K}_{12}^{\rm e}=K_{12}^{\rm e}/(k_{\rm B}T)$ (blue) as a function of the phase difference $\phi_2$.
(b) The dimensionless effective odd elastic constant $\hat{k}^{\rm o}=k^{\rm o}/(k_{\rm B}T)$ as a 
function of $\phi_2$.
For both (a) and (b), the other parameters are the same as in Fig.~\ref{fig:rawdata}.
}
\label{fig:KeKophi}
\end{figure}

\begin{figure}[tb]
\centering
\includegraphics[scale=0.45]{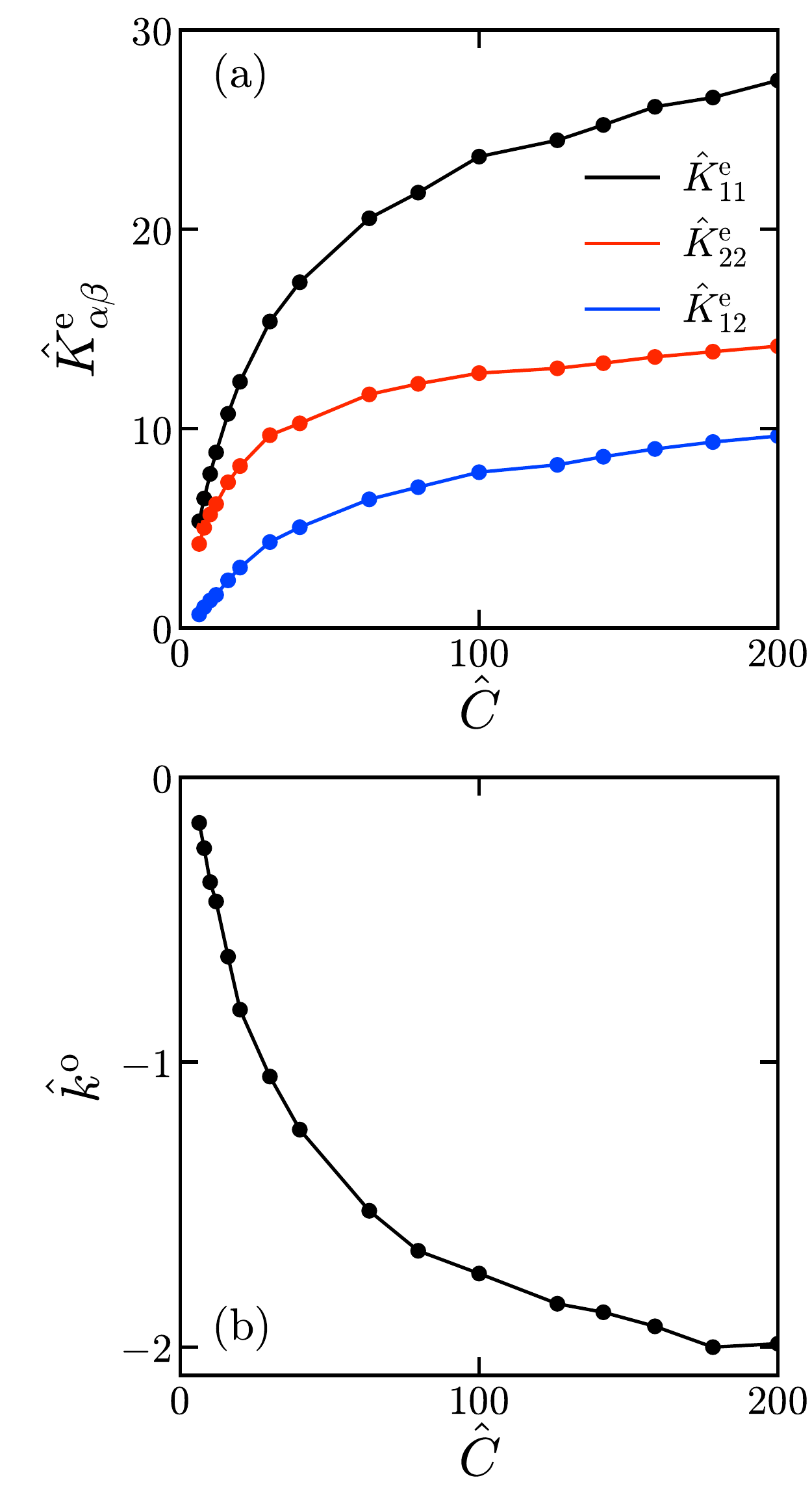}
\caption{(Color online)
(a) The dimensionless effective even elastic constants 
$\hat{K}_{11}^{\rm e}$ (black), $\hat{K}_{22}^{\rm e}$ (red), and $\hat{K}_{12}^{\rm e}$ (blue) as 
a function of the dimensionless coupling  parameter $\hat{C}=C/(k_{\rm B}T)$.
(b) The dimensionless effective odd elastic constant $\hat{k}^{\rm o}$ as a function 
of $\hat{C}$.
For both (a) and (b), the other parameters are the same as in Fig.~\ref{fig:rawdata}.
}
\label{fig:KeKoC}
\end{figure}

\subsection{Effective even and odd elastic constants}

The short-time behaviors of the obtained correlation functions have the following characteristic features; 
(i) the initial slopes of the self-correlation functions $\phi_{11}(t)$ and $\phi_{22}(t)$ are the same, 
(ii) the initial slope of the symmetric part of the cross correlation function $\phi_{12}^\mathrm{S}(t)$
vanishes, i.e., $\dot{\phi}_{12}^\mathrm{S}(0)=0$,
(iii) the initial value of the antisymmetric part of the cross correlation function 
$\bar{\phi}_{12}^\mathrm{A}$ vanishes, i.e., $\bar{\phi}_{12}=\bar{\phi}_{21}$.

Under these conditions, the obtained correlation functions can be interpreted in terms of the following  
coupled odd Langevin equations: 
\begin{align}
\dot x_1&=-\mu(K_{11}x_1+K_{12}x_2)+\zeta_1, 
\label{simpleOverLanEq1}
\\
\dot x_2&=-\mu(K_{21}x_2+K_{22}x_2)+\zeta_2,
\label{simpleOverLanEq2}
\end{align}
where $\mu$ is the effective mobility and $\zeta_\alpha$ ($\alpha=1,2$) is Gaussian white noise
that satisfies the fluctuation dissipation relation.
More general situations and the statistical property of the noises were discussed in Ref.~\cite{YIKLSHK22}.
The effective elastic constant matrix $K_{\alpha\beta}$ can be written in terms 
of even and odd elastic constants as follows [see Eqs.~(\ref{EC-o}) and (\ref{SymK})] 
\begin{align}
\begin{pmatrix}
	K_{11} & K_{12}
	\\
	K_{21} & K_{22} 
\end{pmatrix}
=
\begin{pmatrix}
	K_{11}^{\rm e} & K_{12}^{\rm e} + k^{\rm o}
	\\
	K_{12}^{\rm e} - k^{\rm o}& K_{22}^{\rm e} 
\end{pmatrix},
\label{evenplusodd}
\end{align}
where $K_{11}^{\rm e}$, $K_{12}^{\rm e}$, and $K_{22}^{\rm e}$ are even elastic constants and
$k^{\rm o}$ is the odd elastic constant~\cite{Scheibner20,YHSK21,YKLHSK22}.

In Appendix~\ref{appa}, we show that the short-time behaviors of the symmetric and antisymmetric
parts of the correlation functions obtained from Eqs.~(\ref{simpleOverLanEq1}) and (\ref{simpleOverLanEq2})
are given by [see Eqs.~(\ref{phiSsimple}) and (\ref{phiAsimple})]
\begin{align}
&\phi_{\alpha\beta}^\mathrm S(t)
\approx\bar \phi_{\alpha\beta}- k_\mathrm BT \mu|t| \delta_{\alpha\beta},
\label{textphiSsimple} \\
&\phi_{\alpha\beta}^\mathrm A(t)\approx- 
\frac{2 k^\mathrm o k_\mathrm BT \mu t}{\mathrm{tr}[K^\mathrm e]} \epsilon_{\alpha\beta},
\label{textphiAsimple}
\end{align}
where $\mathrm{tr}[K^\mathrm e]$ is the trace of the matrix $K^{\rm e}_{\alpha\beta}$ and 
$\epsilon_{\alpha\beta}$ is the 2D Levi-Civita tensor.
The equal-time-correlation function $\bar \phi_{\alpha\beta}$ in Eq.~(\ref{textphiSsimple}) 
is now given by [see Eq.~(\ref{app:ETCF-ODLEnewsimple})]
\begin{align}
\bar \phi_{\alpha\beta}& = \frac{k_\mathrm BT}{1+\nu^2} \left[((K^\mathrm e)^{-1})_{\alpha\beta}
+\frac{2\nu^2}{\mathrm{tr}[K^\mathrm e]}\delta_{\alpha\beta} 
\right. \nonumber \\ 
& \left. 
+\frac{k^\mathrm o}{\det[K^{\rm e}]\mathrm{tr}[K^\mathrm e]}
\left( \epsilon_{\alpha\gamma} K^{\rm e}_{\gamma\beta} 
+\epsilon_{\beta\gamma} K^{\rm e}_{\gamma\alpha} \right) \right],
\label{app:ETCF-ODLEnewsimpletext}
\end{align}
where $\det[K^{\rm e}]$ is the determinant of $K^{\rm e}_{\alpha\beta}$,  
$((K^\mathrm e)^{-1})_{\alpha\beta}$ is the inverse of $K^{\mathrm e}_{\alpha\beta}$,
and $\nu^2=(k^\mathrm o)^2/\det[K^\mathrm e]$.
Using these expressions, we extract the effective even and odd elastic constants of the catalytic micromachine.
We first checked that $\mu/\mu_\theta \approx 1.0$ holds in the simulation, which is consistent with 
the choice of the parameter $\mu_s/\mu_\theta=1$.
Hence we can identify $\mu$ with $\mu_s$ and obtain the effective elastic constants.

In Fig.~\ref{fig:KeKophi}(a), we plot the three dimensionless even elastic constants 
$\hat{K}_{\alpha\beta}^{\rm e}=K_{\alpha\beta}^{\rm e}/(k_{\rm B}T)$ as a function of the phase 
difference $\phi_2$. 
We see that they are periodic function of $\phi_2$. 
The period of $K_{11}^{\rm e}$ and $K_{22}^{\rm e}$ is $\pi$, whereas that of $K_{12}^{\rm e}$
is $2\pi$.
In Fig.~\ref{fig:KeKophi}(b), on the other hand, we plot $\hat{k}^{\rm o}=k^{\rm o}/(k_{\rm B}T)$ as a 
function of $\phi_2$.
Notice that $k^{\rm o}$ can take negative values and roughly approximated as 
$k^{\rm o} \sim - \sin \phi_2$.
The present analysis implies that the effective elasticity (both even and odd) can be obtained only 
by measuring the structural dynamics without knowing any detailed dynamics of the 
chemical reaction variable $\theta$.

In Figs.~\ref{fig:KeKoC}(a) and (b), on the other hand, we plot $\hat{K}_{\alpha\beta}^{\rm e}$ and $\hat{k}^{\rm o}$,
respectively, as a function of the coupling parameter $\hat{C}$ while the phase difference is fixed to 
$\phi_2=\pi/2$.
We see in Fig.~\ref{fig:KeKoC}(a) that the even elastic constants $K_{\alpha\beta}^{\rm e}$ increase with $C$.
Although the sign of the odd elastic constant $k^{\rm o}$ is negative in Fig.~\ref{fig:KeKoC}(b), its magnitude 
also increases when $C$ is made larger.

\subsection{Nonreciprocality and power efficiency}
\label{sec:Nonreciprocality}

\begin{figure}[tb]
\centering
\includegraphics[scale=0.45]{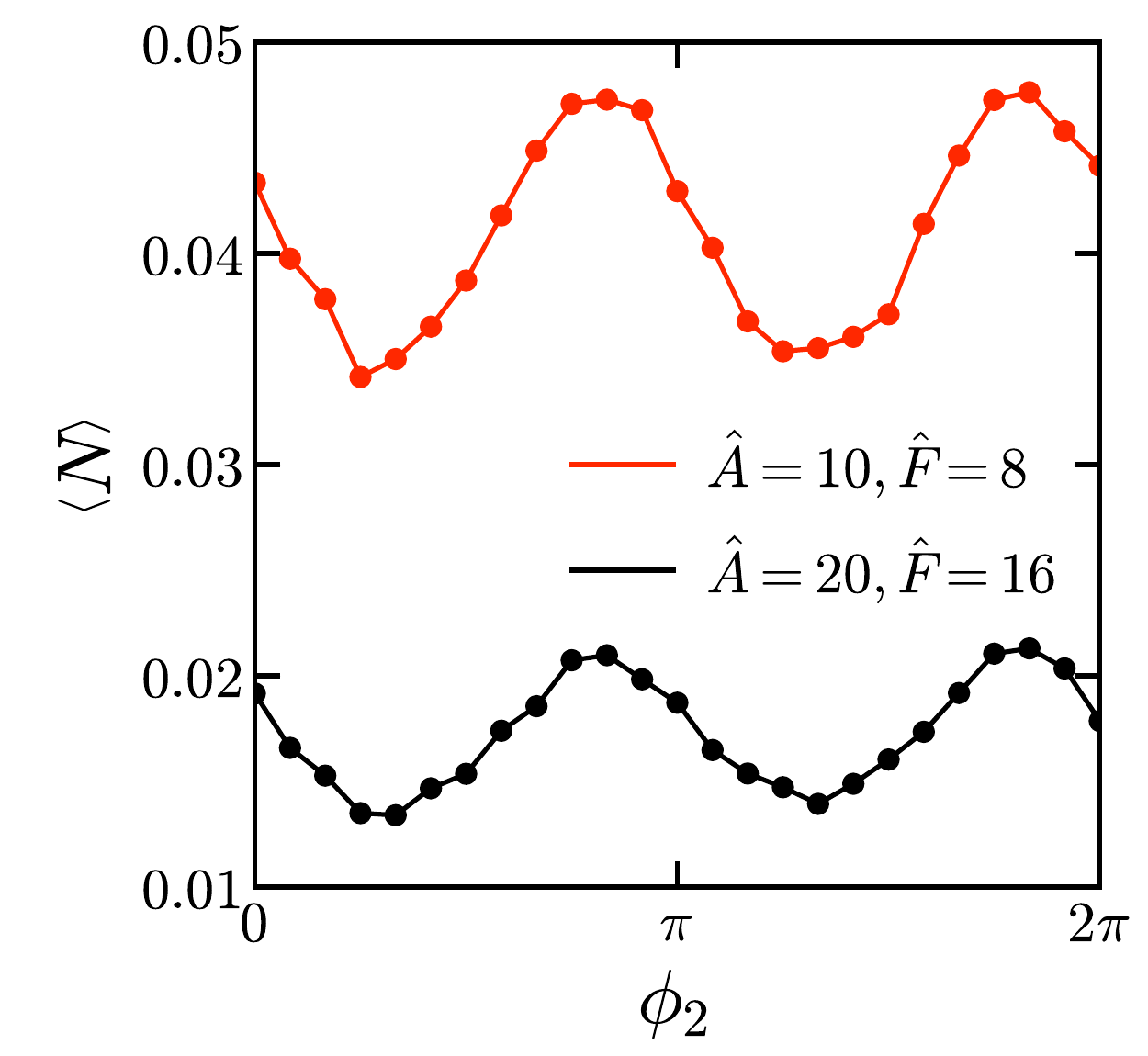}
\caption{(Color online)
The average number of reaction cycle per unit time $\langle N \rangle$ as a function of the phase 
difference $\phi_2$ when $\hat{A}=20$ and $\hat{F}=16$ (black) and when $\hat{A}=10$ and $\hat{F}=8$ (red).
The other parameters are the same as in Fig.~\ref{fig:rawdata}.
}
\label{Fig:rectionspeed}
\end{figure}

\begin{figure}[tb]
\centering
\includegraphics[scale=0.45]{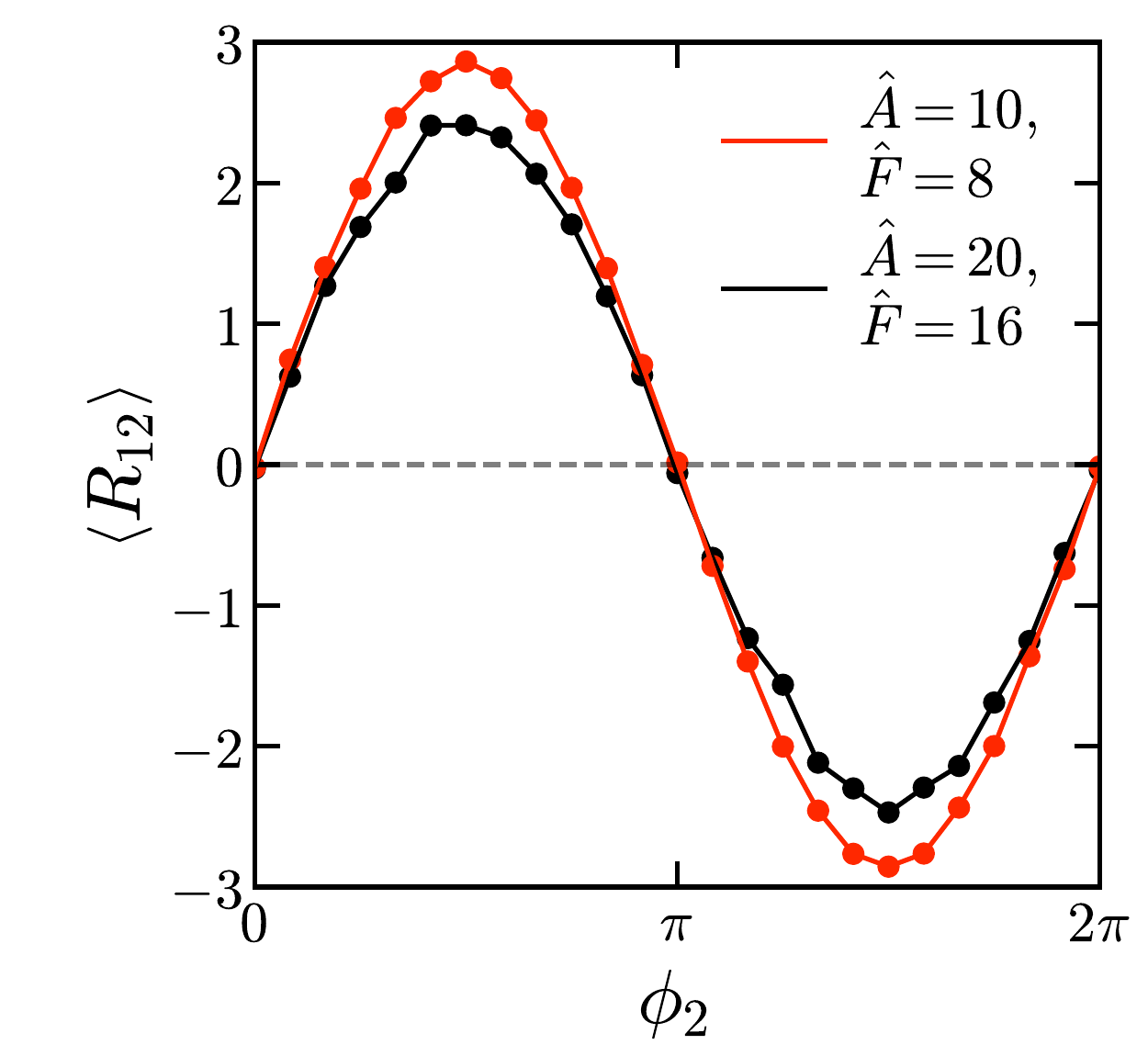}
\caption{(Color online)
The average nonreciprocality $\langle R_{12} \rangle$ as a function of $\phi_2$ when 
$\hat{A}=20$ and $\hat{F}=16$ (black) and when $\hat{A}=10$ and $\hat{F}=8$ (red). 
The other parameters are the same as in Fig.~\ref{fig:rawdata}.
}
\label{Fig:nonreciprocality}
\end{figure}

\begin{figure}[tb]
\centering
\includegraphics[scale=0.45]{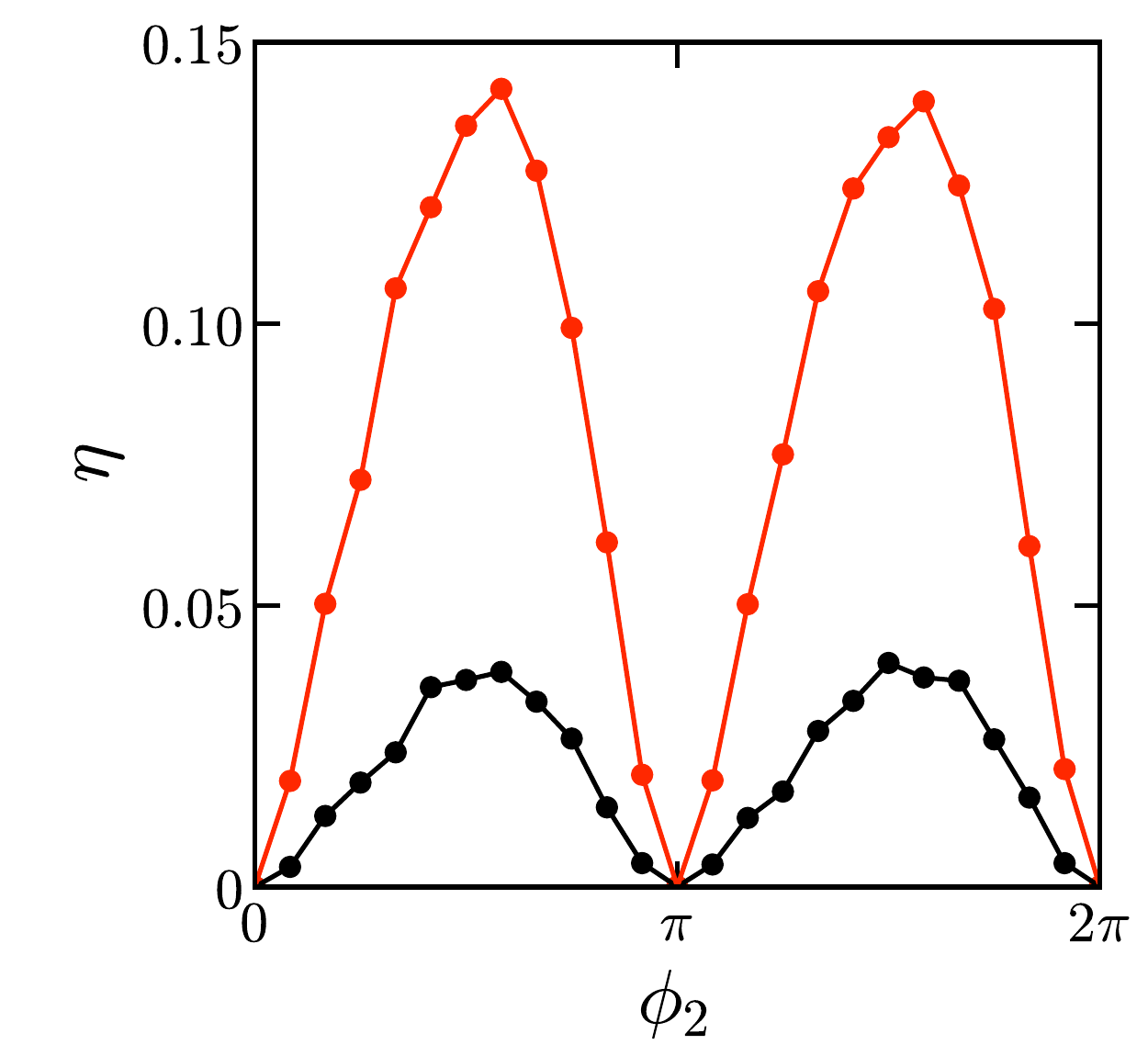}
\caption{(Color online)
The power efficiency $\eta=\langle W \rangle/(2\pi F)$ as a function of the phase difference $\phi_2$
when $\hat{A}=20$ and $\hat{F}=16$ (black) and when $\hat{A}=10$ and $\hat{F}=8$ (red).
The other parameters are the same as in Fig.~\ref{fig:rawdata}.
}
\label{Fig:power}
\end{figure}

To characterize the nonequilibrium degree of a micromachine, it is useful to consider the 
nonreciprocality defined in Eq.~(\ref{nonreciprocality}).
For the current stochastic model, we first discuss the average number of reaction cycle 
$\langle N \rangle$ per unit time that is obtained by dividing the total number of cycles by the 
total simulation time.
In Fig.~\ref{Fig:rectionspeed}, we plot $\langle N \rangle$ as a function of the phase difference 
$\phi_2$ for two combinations of $A$ and $F$ values.
We find that $\langle N \rangle$ is a periodic function of $\phi_2$ and the period is $\pi$.

Next, we calculate the statistical average of nonreciprocality $R_{12}$ over many cycles of chemical reaction.
To do so, we accumulate the quantity $\dot{s}_1 s_2$ over long time and divide it by the total number 
of cycles during the simulation to obtain $\langle R_{12} \rangle$ per one cycle which is plotted 
in Fig.~\ref{Fig:nonreciprocality} as a function of $\phi_2$.
We see that the behavior of $\langle R_{12} \rangle $ is similar to that of $k^{\rm o}$ 
in Fig.~\ref{fig:KeKophi}(b) and can be approximated as $\langle R_{12} \rangle \sim \sin \phi_2$.
This result is reasonable because it was shown before that $\langle R_{12} \rangle$ is given by the 
time derivative of Eq.~(\ref{textphiAsimple}) and can be written as~\cite{Ishimoto22}  
\begin{align}
\langle R_{12} \rangle = -\frac{2  k^\mathrm o k_\mathrm BT  \mu}
{\mathrm{tr}[K^\mathrm e] \langle N \rangle},
\label{R12odd}
\end{align}
where $\langle N \rangle^{-1}$ is the average period of the reaction cycle.
This expression confirms that $\langle R_{12} \rangle$ is proportional to $k^{\rm o}$.
Hence the average nonreciprocality $\langle R_{12} \rangle$ 
becomes nonzero only if there is finite odd elasticity.
We note again that such dynamics occur when the elastic constants are nonreciprocal, 
i.e., $K_{12}\neq K_{21}$ in Eq.~(\ref{evenplusodd}).

The presence of odd elasticity indicates that mechanical work is done by the micromachine through its 
structural change and the average work per unit cycle is given by~\cite{Scheibner20}   
\begin{align}
\langle W \rangle = -2 k^{\rm o} \langle R_{12} \rangle.
\label{work_cycle}
\end{align}
In Fig.~\ref{Fig:power}, we plot the average power efficiency defined by $\eta=\langle W \rangle/(2\pi F)$ 
as a function of the phase difference $\phi_2$ by using the simulation result.
Notice that the energy supplied during one cycle of the catalytic reaction is $2\pi F$.
The power efficiency vanishes when $\phi_2=0$, $\pi$, and $2\pi$ for which the deformation of the 
micromachine is reciprocal. 
This is also reasonable because if we substitute Eq.~(\ref{R12odd}) into Eq.~(\ref{work_cycle}) the average 
work should approximately behave as~\cite{YHSK21} 
\begin{align}
\langle W \rangle \sim (k^\mathrm o)^2 \sim (\sin \phi_2)^2.
\label{scaling}
\end{align}
We also find that the power efficiency increases when the activation energy $A$ and the 
nonequilibrium driving force $F$ become smaller.  
Hence the micromachine can have a higher $\eta$ value when thermal fluctuations are larger compared 
with the potential barrier.

\section{Summary and discussion}
\label{Sec:Dis}

In this paper, we have performed numerical simulations of a model micromachine driven by catalytic 
chemical reactions as previously proposed by some of the present authors~\cite{Yasuda21a}. 
We have obtained the structural time-correlation functions and also extracted their symmetric and 
antisymmetric parts (see Fig.~\ref{fig:timecorrelation}).
These results were further analyzed in terms of odd Langevin dynamics and we have obtained 
the effective even and odd elastic constants of the catalytic system (see Figs.~\ref{fig:KeKophi} and 
\ref{fig:KeKoC}). 
The presence of the odd elasticity demonstrates the broken time-reversal symmetry and we have 
confirmed that the nonreciprocality $\langle R_{12} \rangle$ is proportional to the odd elastic 
constant $k^\mathrm o$ (see Fig.~\ref{Fig:nonreciprocality}).
We have also calculated the power efficiency of a micromachine and found that it increases when 
the activation energy $A$ becomes smaller (see Fig.~\ref{Fig:power}).

Let us summarize here our perspectives on stochastic micromachines, which have been discussed 
in our previous papers~\cite{Yasuda21a,YIKLSHK22} as well as in this work that completes our view.
First, the concept of odd elasticity is not necessarily limited to elastic materials as in the original 
work~\cite{Scheibner20} but can also be applied to general nonreciprocal systems with odd interactions. 
Second, we have explicitly discussed the role of thermal fluctuations in catalytic micromachines and 
showed that the average work is proportional to the square of the odd elastic constant [see
Eq.~(\ref{scaling})]. 
Third, we have suggested how to extract nonequilibrium properties of a micromachine only by measuring 
its structural dynamics even when the chemical reaction variable is a hidden nonequilibrium variable.

Recently, we have derived dynamical equations for a nonequilibrium active system with odd elasticity
by using Onsager's variational principle~\cite{LYIHK22}.
We showed that the elimination of an extra variable that is coupled to the nonequilibrium driving 
force leads to the nonreciprocal set of equations for the structural coordinates~\cite{Fruchart21}. 
The obtained nonreciprocal equations manifest the physical origin of the odd elastic constants 
that are proportional to the nonequilibrium force and the friction coefficients~\cite{Scheibner20}.
In the present work, the reaction variable $\theta(t)$ is driven by the nonequilibrium force $F$
and coupled to the structural variables $s_i$.  
Similar to Ref.~\cite{LYIHK22}, the elimination of the nonequilibrium variable results in the 
nonreciprocal deformation of a micromachine characterized by $R_{12}$ in Eq.~(\ref{nonreciprocality}).
This explains how a micromachine converts chemical energy to mechanical work and such a mechanism 
can be common for proteins, enzymes, and even microswimmers~\cite{YHSK21,KYLSHK23}.

We have analyzed the time-correlation functions of a catalytic micromachine in terms of overdamped 
Langevin equations with odd elasticity as shown in Eq.~(\ref{OverLanEq}). 
In Appendix~\ref{appa}, we have assumed that the corresponding mobility matrix $M_{\alpha\beta}$ is 
symmetric, i.e., $M_{\alpha\beta}=M_{\beta\alpha}$, according to Onsager's reciprocal relation~\cite{DoiBook}.
If we remove this assumption and if $M_{\alpha\beta}$ is not symmetric, one also needs to 
consider the odd viscosity~\cite{Avron98}.
Odd viscosity accounts for the fluid flow perpendicular to the velocity gradient and does not contribute 
to energy dissipation~\cite{Banerjee17,Hosaka21,Hosaka21b,Hosaka23}.
Hence, odd viscosity needs to be taken into account when the surrounding environment of a 
micromachine is in out-of-equilibrium situations. 
In such cases, however, we cannot rely on the fluctuation dissipation relation that we have used in 
the derivation of the correlation functions~\cite{YIKLSHK22}.

Finally, we discuss some typical values of the parameters used in our numerical simulations and 
evaluate the effective odd elastic constant. 
According to the experiments on F$_1$-ATPase, the activation energy and the chemical potential 
difference can be roughly estimated as $A \approx 10$\,$k_{\rm B}T$~\cite{Hayashi15} and 
$F \approx 3$\,$k_{\rm B}T$~\cite{Toyabe10}.
Using the relation for work per cycle $\langle W \rangle = 2\pi F \eta$ as in the previous section, 
we estimate $\langle W \rangle \approx 2$\,$k_{\rm B}T$ when $\eta \approx 0.1$ (see 
Fig.~\ref{Fig:power}).
We shall further identify this work as $\langle W \rangle \sim \kappa^{\rm o} d^2$, 
where $\kappa^{\rm o}$ is the odd elastic constant in unit of J/m$^2$ ($k^{\rm o}$ in this paper 
has the dimension of energy) and $d \approx 10^{-8}$\,m is the typical enzyme size~\cite{YIKLSHK22}.
Then one can roughly estimate as $\kappa^{\rm o} \approx10^{-4}$\,J/m$^2$, which is also 
consistent with our previous estimate for a kinesin molecule~\cite{YIKLSHK22}.

\begin{acknowledgments}

K.Y.\ acknowledges the support by a Grant-in-Aid for JSPS Fellows (No.\ 22KJ1640) from the 
Japan Society for the Promotion of Science (JSPS).
K.I.\ acknowledges the JSPS, KAKENHI for Transformative Research Areas A (No.\ 21H05309) 
and the Japan Science and Technology Agency (JST), PRESTO Grant (No.\ JPMJPR1921).
K.Y.\ and K.I.\ were supported by the Research Institute for Mathematical Sciences, an International 
Joint Usage/Research Center located in Kyoto University.
S.K.\ acknowledges the support by the National Natural Science Foundation of China (Nos.\ 12274098 and 
12250710127) and the startup grant of Wenzhou Institute, University of Chinese Academy of Sciences 
(No.\ WIUCASQD2021041).
A.K.\ and K.Y. contributed equally to this work.
\end{acknowledgments}

\appendix
\begin{widetext}

\section{Time-correlation functions for odd Langevin systems}
\label{appa}

In this Appendix, we briefly review the results of the time-correlation functions for odd Langevin 
systems~\cite{YIKLSHK22}.
As shown in Fig.~\ref{Fig:mod}(a), we consider a deformable object such as an enzyme in a passive viscous fluid.
We investigate its dynamics driven by the energy injection owing to catalytic chemical reactions.
For this purpose, we consider an overdamped Langevin system with an odd elastic tensor. 
The Langevin equation for the state variables $x_\alpha(t)$ ($\alpha=1,2, 3, \dots$) can be written 
as~\cite{KuboBook,DoiBook,Weiss03,Weiss07,Weiss20}
\begin{align}
\dot x_\alpha=-M_{\alpha\beta}K_{\beta\gamma}x_\gamma+\zeta_\alpha,
\label{OverLanEq}
\end{align}
where $M_{\alpha\beta}$ is the mobility tensor that is symmetric, i.e., 
$M_{\alpha\beta}=M_{\beta\alpha}$, due to Onsager's reciprocal relation~\cite{KuboBook,DoiBook}.
Moreover, $M_{\alpha\beta}$ is positive definite according to the second law of thermodynamics.
In Eq.~(\ref{OverLanEq}), $\zeta_\alpha$ is Gaussian white noise that has following statistical properties:
\begin{align}
&\langle\zeta_\alpha(t)\rangle=0, \quad \langle \zeta_\alpha(t)\zeta_\beta(t')\rangle =2 D_{\alpha\beta}\delta(t-t'),
\end{align}
where $D_{\alpha\beta}$ is the diffusion tensor and satisfies the fluctuation dissipation relation
$D_{\alpha\beta}=k_{\rm B}T M_{\alpha\beta}$~\cite{KuboBook,DoiBook}.
The equal-time-correlation function $\bar\phi_{\alpha\beta}=\langle x_\alpha x_\beta \rangle$ obeys the Lyapunov equation~\cite{Weiss03,Weiss07,Weiss20}: 
\begin{align}
M_{\alpha\gamma}K_{\gamma\delta}\bar\phi_{\delta\beta}+M_{\beta\gamma}K_{\gamma\delta}\bar\phi_{\delta\alpha}=2D_{\alpha\beta}.
\end{align}

In Eq.~(\ref{OverLanEq}), $K_{\alpha\beta}$ is the elastic constant tensor.
For active systems with nonconservative interactions, $K_{\alpha\beta}$ can have an antisymmetric 
part that corresponds to the odd elasticity~\cite{Scheibner20,YHSK21,YKLHSK22}.
Hence, $K_{\alpha\beta}$ can generally be written as 
\begin{align}
K_{\alpha\beta}=K_{\alpha\beta}^{\rm e}+K_{\alpha\beta}^{\rm o},
\label{EC-o}
\end{align}
where the symmetric (even) part and the antisymmetric (odd) part satisfy 
$K_{\alpha\beta}^{\rm e}=K_{\beta\alpha}^{\rm e}$ 
and  $K_{\alpha\beta}^{\rm o}=-K_{\beta\alpha}^{\rm o}$, respectively.
We further consider a system with only two degrees of freedom ($\alpha, \beta=1,2$) and assume that 
the elastic tensor is given by the following form:
\begin{align}
K_{\alpha\beta}=K_{\alpha\beta}^\mathrm e + k^\mathrm o \epsilon_{\alpha\beta},
\label{SymK}
\end{align} 
where $k^\mathrm o$ is the scaler odd elastic constant and $\epsilon_{\alpha\beta}$ is the 2D Levi-Civita 
tensor with $\epsilon_{11}=\epsilon_{22}=0$ and $\epsilon_{12}=-\epsilon_{21}=1$.

In our previous paper~\cite{YIKLSHK22}, we showed that the equal-time-correlation function when 
$\alpha, \beta=1,2$ becomes 
\begin{align}
\bar \phi_{\alpha\beta}&=\frac{k_\mathrm BT}{1+\nu^2}\left[((K^\mathrm e)^{-1})_{\alpha\beta}+\frac{2\nu^2}{\mathrm{tr}[M K^\mathrm e]}M_{\alpha\beta}-\frac{k^\mathrm o\det[M]}{\mathrm{tr}[M K^\mathrm e]}\left[\epsilon_{\alpha\gamma}(M^{-1})_{\gamma\delta} ((K^\mathrm e)^{-1})_{\delta\beta}+\epsilon_{\beta\gamma}(M^{-1})_{\gamma\delta} ((K^\mathrm e)^{-1})_{\delta\alpha}\right]\right],
\label{app:ETCF-ODLE}
\end{align}
where ``tr" and ``det" indicate the trace and determinant of the matrix, respectively, 
$((K^\mathrm e)^{-1})_{\alpha\beta}$ is the inverse of $K^{\mathrm e}_{\alpha\beta}$,
and $\nu^2=(k^\mathrm o)^2/\det[K^\mathrm e]$  (see Eq.~(C2) in Ref.~\cite{YIKLSHK22}).
The above expression can be rewritten as 
\begin{align}
\bar \phi_{\alpha\beta}&=\frac{k_\mathrm BT}{1+\nu^2}\left[((K^\mathrm e)^{-1})_{\alpha\beta}+\frac{2\nu^2}{\mathrm{tr}[M K^\mathrm e]}M_{\alpha\beta}
+\frac{k^\mathrm o}{\det[K^{\rm e}]\mathrm{tr}[M K^\mathrm e]}
\left[\epsilon_{\alpha\gamma} K^{\rm e}_{\gamma\delta} M_{\delta\beta}
+\epsilon_{\beta\gamma} K^{\rm e}_{\gamma\delta} M_{\delta\alpha}\right]\right].
\label{app:ETCF-ODLEnew}
\end{align}
In the short-time limit, the time-correlation function can be decomposed into the symmetric and antisymmetric 
parts as $\phi_{\alpha\beta}(t)=\phi_{\alpha\beta}^\mathrm S(t) + \phi_{\alpha\beta}^\mathrm A(t)$, and 
they are respectively given by 
\begin{align}
&\phi_{\alpha\beta}^\mathrm S(t)
\approx\bar \phi_{\alpha\beta}- k_\mathrm BT M_{\alpha\beta}|t|,
\label{phiS} \\
&\phi_{\alpha\beta}^\mathrm A(t)\approx- 
\frac{2 k^\mathrm o k_\mathrm BT \det[M] t}{\mathrm{tr}[M K^\mathrm e]} \epsilon_{\alpha\beta},
\label{phiA}
\end{align}
(see Eq.~(C6) in Ref.~\cite{YIKLSHK22}).

According to our simulation results, the mobility matrix can be further simplified as 
$M_{\alpha\beta}=\mu \delta_{\alpha\beta}$. 
In this case, Eqs.~(\ref{app:ETCF-ODLEnew}), (\ref{phiS}), and (\ref{phiA}) further become
\begin{align}
\bar \phi_{\alpha\beta}&=\frac{k_\mathrm BT}{1+\nu^2}\left[((K^\mathrm e)^{-1})_{\alpha\beta}
+\frac{2\nu^2}{\mathrm{tr}[K^\mathrm e]}\delta_{\alpha\beta}
+\frac{k^\mathrm o}{\det[K^{\rm e}]\mathrm{tr}[K^\mathrm e]}
\left( \epsilon_{\alpha\gamma} K^{\rm e}_{\gamma\beta} 
+\epsilon_{\beta\gamma} K^{\rm e}_{\gamma\alpha} \right) \right],
\label{app:ETCF-ODLEnewsimple}
\end{align}
\begin{align}
&\phi_{\alpha\beta}^\mathrm S(t)
\approx\bar \phi_{\alpha\beta}-  k_\mathrm BT \mu|t| \delta_{\alpha\beta},
\label{phiSsimple} \\
&\phi_{\alpha\beta}^\mathrm A(t)\approx- 
\frac{2 k^\mathrm o k_\mathrm BT  \mu t}{\mathrm{tr}[K^\mathrm e]} \epsilon_{\alpha\beta}.
\label{phiAsimple}
\end{align}
These expression are used to extract the effective elastic constants of a catalytic micromachine.

\end{widetext}


\begin{thebibliography}{99}


\bibitem{Toyabe15}
S. Toyabe and M. Sano, 
J. Phys. Soc. Jpn. 84, 102001 (2015). 

\bibitem{Bechinger16}
C. Bechinger, R. Di Leonardo, H. L\"owen, C. Reichhardt, G. Volpe, and G. Volpe, 
Rev. Mod. Phys. 88, 045006 (2016).

\bibitem{Brown20}
A. I. Brown and D. A. Sivak, 
Chem. Rev. 120, 434 (2020). 

\bibitem{Dey16}
K. K. Dey, F. Wong, A. Altemose, and A. Sen,
Curr. Opin. Colloid Interface Sci. 21, 4 (2016).

\bibitem{HK22}
Y. Hosaka and S. Komura, 
Biophys. Rev. Lett. 17, 51 (2022).

\bibitem{Togashi10}
Y. Togashi, T. Yanagida, and A. S. Mikhailov, 
PLoS Comput. Biol. 6, e1000814 (2010).

\bibitem{Mugnai20}
M. L. Mugnai, C. Hyeon, M. Hinczewski, and D. Thirumalai, 
Rev. Mod. Phys. 92, 025001 (2020).

\bibitem{Hosaka20}
Y. Hosaka, S. Komura, and A. S. Mikhailov,
Soft Matter 16, 10734 (2020).

\bibitem{Harada05}
T. Harada and S.-i. Sasa, 
Phys. Rev. Lett. 95, 130602 (2005). 

\bibitem{Toyabe10}
S. Toyabe, T. Okamoto, T. Watanabe-Nakayama, H. Taketani, S. Kudo, and E. Muneyuki, 
Phys. Rev. Lett. 104, 198103 (2010).

\bibitem{Hayashi15}
R. Hayashi, K. Sasaki, S. Nakamura, S. Kudo, Y. Inoue, H. Noji, and K. Hayashi, 
Phys. Rev. Lett. 114, 248101 (2015).

\bibitem{Ariga18}
T. Ariga, M. Tomishige, and D. Mizuno, 
Phys. Rev. Lett. 121, 218101 (2018).

\bibitem{Yasuda21a}
K. Yasuda and S. Komura, 
Phys. Rev. E 103, 062113 (2021).

\bibitem{Golestanian08}
R. Golestanian and A. Ajdari, 
Phys. Rev. E \textbf{77}, 036308 (2008).

\bibitem{Sou19}
I. Sou, Y. Hosaka, K. Yasuda, and S. Komura, 
Phys. Rev. E \textbf{100}, 022607 (2019).

\bibitem{Sou21}
I. Sou, Y. Hosaka, K. Yasuda, and S. Komura, 
Physica A \textbf{562}, 125277
(2021).

\bibitem{Leoni17}
M. Leoni and P. Sens, 
Phys. Rev. Lett. \textbf{118}, 228101 (2017).

\bibitem{Tarama18}
M. Tarama and R. Yamamoto, 
J. Phys. Soc. Jpn. \textbf{87}, 044803 (2018).

\bibitem{Scheibner20}
C. Scheibner, A. Souslov, D. Banerjee, P. Sur\'{o}wka, W. T. Irvine, and V. Vitelli,
Nat. Phys. 16, 475 (2020).

\bibitem{Fruchart22}
M. Fruchart, C. Scheibner, and V. Vitelli, 
Annu. Rev. Condens. Matter Phys. 14, 471 (2023).

\bibitem{Zhou20}
D. Zhou and J. Zhang,
Phys. Rev. Res. 2, 023173 (2020).

\bibitem{Braverman21}
L. Braverman, C. Scheibner, B. VanSaders, and V. Vitelli, 
Phys. Rev. Lett. 127, 268001 (2021).

\bibitem{Tan22}
T. H. Tan, A. Mietke, J. Li, Y. Chen, H. Higinbotham, P. J. Foster, S. Gokhale, J. Dunkel, and N. Fakhri,
Nature 607, 287 (2022).

\bibitem{YHSK21} 
K. Yasuda, Y. Hosaka, I. Sou, and S. Komura,
J. Phys. Soc. Jpn. 90, 075001 (2021).

\bibitem{KYLSHK23} 
A. Kobayashi, K. Yasuda, L.-S. Lin, I. Sou, Y. Hosaka, and S. Komura,
J. Phys. Soc. Jpn. 92, 034803 (2023).

\bibitem{Ishimoto22} 
K. Ishimoto, C. Moreau, and K. Yasuda, 
Phys. Rev. E 105, 064603 (2022). 

\bibitem{Brandenbourger22}
M. Brandenbourger, C. Scheibner, J. Veenstra, V. Vitelli, and C. Coulais,
arXiv:2108.08837

\bibitem{YIKLSHK22}
K. Yasuda, K. Ishimoto, A. Kobayashi, L.-S. Lin, I. Sou, Y. Hosaka, and S. Komura,
J. Chem. Phys. 157, 095101 (2022).

\bibitem{YKLHSK22}
K. Yasuda, A. Kobayashi, L.-S. Lin, Y. Hosaka, I. Sou, and S. Komura,
J. Phys. Soc. Jpn. 91, 015001 (2022).

\bibitem{LYIHK22}
L.-S. Lin, K. Yasuda, K. Ishimoto, Y. Hosaka, and S. Komura,
J. Phys. Soc. Jpn. 92, 033001 (2023).

\bibitem{Dillbook}
K. Dill and S. Bromberg,
\textit{Molecular Driving Forces: Statistical Thermodynamics in Biology, Chemistry, Physics, and Nanoscience}
(Garland Science, London and New York, 2010).

\bibitem{Hanggi90}
P. H\"anggi, P. Talkner, and M. Borkovec, 
Rev. Mod. Phys. 62, 251 (1990).

\bibitem{Mikhailov15}
A. S. Mikhailov and R. Kapral, 
Proc. Natl. Acad. Sci. USA 112, E3639 (2015).

\bibitem{Canalejo21}
J. Agudo-Canalejo, T. Adeleke-Larodo, P. Illien, and R. Golestanian,
Phys. Rev. Lett. 127, 208103 (2021).

\bibitem{KuboBook}
R. Kubo, M. Toda, and N. Hashitsume, 
\textit{Statistical Physics II} (Springer, New York, 1991).

\bibitem{DoiBook}
M. Doi, 
\textit{Soft Matter Physics}
(Oxford University Press, Oxford, 2013).

\bibitem{Caloz18}
C. Caloz, A. Al\`u, S. Tretyakov, D. Sounas, K. Achouri, and Z.-L. Deck-L\'{e}ger,
Phys. Rev. Applied 10, 047001 (2018).

\bibitem{Nassar20}
H. Nassar, B. Yousefzadeh, R. Fleury, M. Ruzzene, A. Al\`u, C. Daraio, A. N. Norris, G. Huang, M. R. Haberman,
Nat. Rev. Mater. 5, 667 (2020). 

\bibitem{Zhou22}
D. Zhou, D. Z. Rocklin, M. Leamy, and Y. Yao,
Nat. Commun. 13,  3379 (2022).

\bibitem{Epstein20}
J. M. Epstein and K. K. Mandadapu, 
Phys. Rev. E 101, 052614 (2020).

\bibitem{Hargus20}
C. Hargus, K. Klymko, J. M. Epstein, and K. K. Mandadapu,
J. Chem. Phys. 152, 201102 (2020).

\bibitem{Han21}
M. Han, M. Fruchart, C. Scheibner, S. Vaikuntanathan, J. J. de Pablo, and V. Vitelli,
Nat. Phys. 17, 1260 (2021).

\bibitem{Fruchart21}
M. Fruchart, R. Hanai, P. B. Littlewood, and V. Vitelli, 
Nature 592, 363 (2021).\bibitem{Avron98}
J. E. Avron, 
J. Stat. Phys. 92, 543 (1998).

\bibitem{Banerjee17}
D. Banerjee, A. Souslov, A. G. Abanov, and V. Vitelli, 
Nat. Commun. 8, 1573 (2017).

\bibitem{Hosaka21}
Y. Hosaka, S. Komura, and D. Andelman,
Phys. Rev. E 103, 042610 (2021).

\bibitem{Hosaka21b}
Y. Hosaka, S. Komura, and D. Andelman,
Phys. Rev. E 104, 064613 (2021).

\bibitem{Hosaka23}
Y. Hosaka, D. Andelman, and S. Komura, 
Eur. Phys. J. E 46, 18 (2023).

\bibitem{Weiss03}
J. B. Weiss, 
Tellus A 55, 208 (2003).

\bibitem{Weiss07}
J. B. Weiss, 
Phys. Rev. E 76, 061128 (2007).

\bibitem{Weiss20}
J. B. Weiss, B. Fox-Kemper, D. Mandal, A. D. Nelson, and R. K. P. Zia, 
J. Stat. Phys. 179, 1010 (2020).

\end{thebibliography}

\end{document}